\documentclass{article}
\title{Formulation of a constrained system in terms of extended Lagrangian and its local symmetries}
\author{A. A. Deriglazov\footnote{alexei.deriglazov@ufjf.edu.br ~ On leave of
absence from Dept. Math. Phys., Tomsk Polytechnical University,
Tomsk, Russia.}}
\date{Dept. de Matematica, ICE, Universidade Federal de Juiz de Fora,\\
MG, Brazil; \\
and \\
LAFEX - CBPF/MCT, Rio de Janeiro, RJ, Brazil.}
\begin{document}
\maketitle
\large

\begin{abstract}
It is shown that an arbitrary singular Lagrangian theory (with first and second class constraints up
to $N$-th stage presented in the Hamiltonian formulation) can be reformulated as a theory with at
most third-stage constraints. The corresponding Lagrangian $\tilde L$ can be obtained by pure
algebraic methods, its manifest form in terms of quantities of the initial formulation is find. Local
symmetries of $\tilde L$ are obtained in closed form. All the first class constraints of the initial
Lagrangian turn out to be gauge symmetry generators for $\tilde L$.
\end{abstract}

\noindent
%{\bf PAC codes:} \\
%{\bf Keywords:} Constrained systems, Local symmetries, Gauge theories

\section{Introduction}
Conventional way to describe a relativistic theory is to formulate it in terms of a singular Lagrangian.
In turn, analysis of the singular theory can be carried out in a Hamiltonian formalism. In this framework,
possible motions of the singular system are restricted to lie on some surface of a phase space. Algebraic
equations of the surface (constraints) can be revealed in the course of a Dirac procedure, the latter in general
case requires a number of stages. According to the order of appearance, the constraints are called primary,
second-stage, ... , N-th stage constraints. All the constraints, beside the primary ones are called higher-stage
constraints below.
Whenever are appeared, the higher-stage constraints represent perceptible problems for analysis of the theory.
In particular, search for local symmetries of the Lagrangian action, which is main subject of the present work,
turns out to be rather nontrivial issue in a general case [1-5].
So, it may be reasonable to adopt a different approach to the problem. Namely, instead of looking for properties of the initial Lagrangian $L$ (provided
all its constraints are known), we work out an equivalent Lagrangian $\tilde L$, the latter implies more transparent
structure of constraints (in fact, all the higher-stage constraints of the original formulation enter into
$\tilde L$ in a manifest form, see the last term in Eq. (\ref{13}) below). It allows one to find infinitesimal local symmetries of the formulation in closed form, in terms of the constraints of the initial formulation.
For some particular examples, such a kind possibility has been tested in the recent work [6]. Here we develop
the formalism for an arbitrary theory, with first and second class constraints up to $N$-th stage presented in the
original formulation $L$. $\tilde L$ is called an extended Lagrangian, since the corresponding complete Hamiltonian turns
out to be closely related with an extended Hamiltonian of the original
formulation\footnote{By definition, the extended
Hamiltonian is obtained from the complete one by addition of the higher-stage constraints with corresponding
Lagrangian multipliers. It is known [8] that the two formulations are equivalent.}. So, in this work we also clarify
a relation among the complete and the extended Hamiltonian formulations of a given theory.

The work is organized as follows. With the aim to fix our notations, we outline in Section 2 the
Hamiltonization procedure for an arbitrary singular Lagrangian theory.
In Section 3 we formulate pure algebraic recipe for construction of the extended Lagrangian. All the higher-stage
constraints of $L$ appear as secondary constraints for $\tilde L$. Besides, we demonstrate that $\tilde L$ is a theory
with at most third-stage constraints. Then it is proved that $\tilde L$ and $L$ are
equivalent\footnote{Popular physical theories usually do not involve more than third-stage constraints (example of
a theory with third-stage constraints is the membrane, in the formulation with world-volume metric). Our result
can be considered as an explanation of this fact.}.
Since the original and the reconstructed formulations are equivalent,
it is matter of convenience to use one or another of them for description of a theory under
investigation\footnote{Let us point out that the higher stage constraints
usually appear in a covariant form. So one expects manifest covariance of the extended formulation.}.
In Section 4 we demonstrate one of advantages of the extended formulation by finding its complete
irreducible set of local symmetries. Properties of the extended formulation for some particular cases of
original gauge algebra are discussed in the Conclusion.

\section{Initial formulation with higher-stage first and second class constraints}
Let $L(q^A, \dot q^B)$ be Lagrangian of singular theory:
$rank\frac{\partial^2 L}{\partial\dot q^A\partial\dot q^B}=[i]<[A]$, defined on
configuration space $q^A, A=1, 2, \ldots , [A]$. From the beginning, it is convenient to rearrange the  initial
variables in such a way
that the rank minor is placed in the upper left corner. Then one has $q^A=(q^i, q^\alpha)$,
$i=1, 2, \ldots , [i]$, ~
$\alpha=1, 2, \ldots , [\alpha]=[A]-[i]$, where
$\det\frac{\partial^2 L}{\partial\dot q^i\partial\dot q^j}\ne 0$.

Let us construct the Hamiltonian formulation for the theory. To fix our notations, we carry out the
Hamiltonization procedure in some details. One introduces conjugate
momenta according to the equations
$p_i$ $=$ $\frac{\partial L}{\partial\dot q^i}$, $p_\alpha$ $=$ $\frac{\partial L}{\partial\dot q^\alpha}$.
They are considered as algebraic equations for determining velocities $\dot q^A$.
According to the rank condition, the first $[i]$ equations
can be resolved with respect to $\dot q^i$, let us denote the solution as
\begin{eqnarray}\label{2}
\dot q^i=v^i(q^A, p_j, \dot q^\alpha).
\end{eqnarray}
It can be substituted into remaining $[\alpha]$ equations for the momenta. By construction, the
resulting expressions do not depend on $\dot q^A$ and are called primary constraints
$\Phi_\alpha(q, p)$ of the Hamiltonian formulation. One finds
\begin{eqnarray}\label{3}
\Phi_\alpha\equiv p_\alpha-f_\alpha(q^A, p_j)=0,
\end{eqnarray}
where
\begin{eqnarray}\label{4}
f_\alpha(q^A, p_j)\equiv\left.\frac{\partial L}{\partial\dot q^\alpha}
\right|_{\dot q^i=v^i(q^A, p_j, \dot q^\alpha)}.
\end{eqnarray}
The original equations for the momenta are thus equivalent to the system (\ref{2}), (\ref{3}).
By construction, one has the identities
\begin{eqnarray}\label{1}
\left.\frac{\partial L(q, \dot q)}{\partial\dot q^i}\right|_{\dot q^i\rightarrow v^i(q^A, p_j, \dot q^\alpha)}\equiv p_i, \qquad
\left.v^i(q^A, p_j, \dot q^\alpha)\right|_{p_j\rightarrow\frac{\partial L}{\partial\dot q^j}}\equiv\dot q^i.
\end{eqnarray}

Next step of the Hamiltonian procedure is to introduce an extended phase space parameterized by the
coordinates $q^A, p_A, v^\alpha$, and to define a complete Hamiltonian $H$ according to the rule
\begin{eqnarray}\label{5}
H(q^A, p_A, v^\alpha)=H_0(q^A, p_j)+v^\alpha\Phi_\alpha(q^A, p_B),
\end{eqnarray}
where
\begin{eqnarray}\label{6}
H_0=\left.(p_i\dot q^i-L+ \dot q^\alpha\frac{\partial L}{\partial \dot q^\alpha})\right|
_{\dot q^i\rightarrow v^i(q^A, p_j, \dot q^\alpha)}.
\end{eqnarray}
Then the following system of equations on this space
\begin{eqnarray}\label{7}
\dot q^A=\{q^A, H\}, \qquad \dot p_A=\{p_A, H\}, \qquad
\Phi_\alpha(q^A, p_B)=0,
\end{eqnarray}
is equivalent to the Lagrangian equations following from $L$, see [8]. Here $\{ , \}$ denotes
the Poisson bracket.

From Eq. (\ref{7}) it follows that all the solutions are confined to lie on a surface of the extended phase space
defined by the algebraic equations $\Phi_\alpha=0$. It may happen, that
the system (\ref{7}) contains in reality more then $[\alpha]$ algebraic equations. Actually, derivative of the
primary constraints with respect to time implies, as algebraic consequences of the system (\ref{7}),
the so called second stage equations: $\{\Phi_\alpha, H\}$ $\equiv$
$\{\Phi_{\alpha}, \Phi_\beta\}v^\beta+\{\Phi_{\alpha}, H_0\}$ $=$ $0$.
They can be added to Eq. (\ref{7}), which gives an equivalent system.
Let on-shell one has $rank\{\Phi_{\alpha}, \Phi_\beta\}=[\alpha ']\leq [\alpha]$. Then
$[\alpha ']$
equations of the second-stage system
can be used to represent some $v^{\alpha '}$ through other variables. It can be substituted into the remaining
$[\alpha '']\equiv [\alpha]-[\alpha ']$ equations, the resulting expressions do not contain $v^\alpha$ at all.
Thus the second-stage system can be presented in the equivalent form
\begin{eqnarray}\label{7.0}
v^{\alpha '}=v^{\alpha '}(q^A, p_j, v^{\alpha ''}), \qquad T_{\alpha ''}(q^A, p_j)=0.
\end{eqnarray}
Functionally independent equations among $T_{\alpha ''}=0$, if any, represent secondary Dirac constraints.
Thus all the
solutions of the system (\ref{7}) are confined to the surface defined by $\Phi_\alpha=0$ and by the
equations (\ref{7.0}).

The secondary constraints may
imply third-stage constraints, and so on. We suppose that the theory has
constraints up to $N$-th stage, $N\ge 2$. Higher stage constraints are denoted by $T_a(q^A, p_j)=0$.
Then the complete constraint system is
$G_I\equiv(\Phi_\alpha, T_a)$, while  all the solutions of Eq. (\ref{7}) are confined to the
surface defined by the equations $\Phi_\alpha=0$ as well as by\footnote{It is known
[8], that the procedure reveals all the algebraic equations presented in the system (\ref{7}). Besides, surface
of solutions of Eq. (\ref{7}) coincides with the surface $\Phi_\alpha=0$, $\{ G_I, H\}=0$.}
\begin{eqnarray}\label{7.1}
\{ G_I, H\}=0.
\end{eqnarray}
By construction, after substitution of the velocities determined during the Dirac procedure, these equations vanish
on the complete constraint surface $G_J$.

Suppose that $\{G_I, G_J\}=\triangle_{IJ}(q^A, p_j)$, where
$\left. rank\triangle_{IJ}\right|_{G_I=0}=[I_2]<[I]$, that is both first and second class constraints are presented.
It will be convenient to separate them. According to the rank condition, there exist $[I_1]={I}-[I_2]$
independent null-vectors $\vec K_{I_1}$ of the matrix $\triangle$ on the surface $G_I=0$, with the components
$K_{I_1}{}^J(q^A, p_j)$. Then bracket of the constraints $G_{I_1}\equiv K_{I_1}{}^JG_J$ with any $G_I$ vanishes, hence
$G_{I_1}$ represent first class subset. Let $K_{I_2}{}^J(q^A, p_j)$ be any completion of the set
$K_{I_1}{}^J$ up to a basis of $[I]$-dimensional vector space. By construction, the matrix
\begin{eqnarray}\label{7.2}
K_{I}{}^J\equiv
\left(K_{I_1}{}^J\atop  K_{I_2}{}^J\right),
\end{eqnarray}
is invertible. So the system $\tilde G_I$ $\equiv$ $(G_{I_1}\equiv K_{I_1}{}^JG_J$, $G_{I_2}\equiv K_{I_2}{}^JG_J)$
is equivalent to the initial system of constraints $G_I$. The constraints $G_{I_2}$ form the second class
subset of the complete set. In arbitrary theory, the constraints obey the following Poisson bracket algebra:
\begin{eqnarray}\label{8}
\{\tilde G_I, \tilde G_J\}=\triangle_{IJ}(q^A, p_B), \qquad \quad \qquad \qquad \quad \cr
\{G_{I_1}, G_J\}=c_{I_1 J}{}^K(q^A, p_B)G_K, \quad \{G_{I_1}, H_0\}=b_{I_1}{}^J(q^A, p_B)G_J, \cr
\{G_{I_2}, G_{J_2}\}=\triangle_{I_2 J_2}(q^A, p_B), \qquad \qquad \quad \qquad
\end{eqnarray}
where
\begin{eqnarray}\label{8.1}
\left. rank\triangle_{IJ}\right|_{G_I=0}=[I_2], \qquad
\left. \det\triangle_{I_2 J_2}\right|_{G_I=0}\ne 0.
\end{eqnarray}

\section{Construction of the extended Lagrangian and its properties}
Starting from the theory described above, we
construct here a Lagrangian $\tilde L(q^A, \dot q^A, s^a)$ defined on the
configuration space $q^A, s^a$. In the Hamiltonian formalism,
it leads to the Hamiltonian\footnote{Let us stress once again, that in our formulation the
variables $s^a$ represent a part of the configuration-space variables.}
$H_0+s^aT_a$, and to the primary constraints $\Phi_\alpha=0, ~ \pi_a=0$, where $\pi_a$ represent
conjugate momenta for $s^a$. Due to special form of the Hamiltonian, preservation in time of the primary constraints
implies, that all the higher stage constraints $T_a$ of initial theory appear as secondary constraints for
the theory $\tilde L$. Moreover, the Dirac procedure stops on third stage: $\tilde L$ turns out to be a theory
with at most third-stage constraints presented. Besides, we demonstrate that the
formulations $L$ and $\tilde L$ are equivalent.

To construct the extended Lagrangian for $L$, let us consider the following equations for the variables
$q^A,\omega_j, s^a$:
\begin{eqnarray}\label{9}
\dot q^i-v^i(q^A, \omega_j, \dot q^\alpha)-s^a\frac{\partial T_a(q^A, \omega_j)}{\partial\omega_i}=0.
\end{eqnarray}
Here the functions $v^i(q^A, \omega_j, \dot q^\alpha)$, ~ $T_a(q^A, \omega_j)$
are taken from the initial formulation.
The equations can be resolved algebraically with respect to $\omega_i$ in a neighboard of the
point $s^a=0$. Actually, Eq. (\ref{9}) with $s^a=0$ coincides with Eq. (\ref{2}) of the initial formulation,
the latter can be resolved, see Eq. (\ref{1}). Hence
$\det\frac{\partial (Eq.(\ref{9}))^i}{\partial\omega_j}\ne 0$ at the
point $s^a=0$. Then the same is true in some vicinity of this point, and Eq. (\ref{9})
thus can be resolved. Let us denote the solution as
\begin{eqnarray}\label{10}
\omega_i=\omega_i(q^A, \dot q^A, s^a).
\end{eqnarray}
By construction, one has the identities
\begin{eqnarray}\label{11}
\left.\omega_i(q, \dot q, s)\right|_{\dot q^i\rightarrow
v^i(q^A, \omega_j, \dot q^\alpha)+s^a\frac{\partial T_a(q^A, \omega_j)}{\partial\omega_i}}\equiv\omega_i,
\qquad \cr
\left.\left(v^i(q^A, \omega_j, \dot q^\alpha)+s^a\frac{\partial T_a(q^A, \omega_j)}{\partial\omega_i}\right)
\right|_{\omega_i(q, \dot q, s)}\equiv\dot q^i,
\end{eqnarray}
as well as the following property of the function $\omega$
\begin{eqnarray}\label{12}
\left.\omega_i(q^A, \dot q^A, s^a)\right|_{s^a=0}=\frac{\partial L}{\partial\dot q^i}.
\end{eqnarray}
Now, the extended Lagrangian for $L$ is defined according to the expression
\begin{eqnarray}\label{13}
\tilde L(q^A, \dot q^A, s^a)=L(q^A, v^i(q^A, \omega_j, \dot q^\alpha), \dot q^\alpha)+ \cr
\omega_i(\dot q^i-v^i(q^A, \omega_j, \dot q^\alpha))-s^aT_a(q^A, \omega_j), \quad
\end{eqnarray}
where the functions $v^i, \omega_i$ are given by Eqs. (\ref{2}), (\ref{10}).
As compare with the initial Lagrangian, $\tilde L$ involves the new variables $s^a$, in a number
equal to the number of higher stage constraints $T_a$. Let us enumerate some properties
of $\tilde L$
\begin{eqnarray}\label{14}
\tilde L(s^a=0)=L,
\end{eqnarray}
\begin{eqnarray}\label{15}
\left.\frac{\partial\tilde L}{\partial\omega_i}
\right|_{\omega(q, \dot q, s)}=0,
\end{eqnarray}
\begin{eqnarray}\label{16}
\frac{\partial\tilde L}{\partial\dot q^\alpha}=
\left.\left.\frac{\partial L(q^A, v^i, \dot q^\alpha)}{\partial\dot q^\alpha}
\right|_{v^i(q, \omega, \dot q^\alpha)}\right|_{\omega(q, \dot q, s)}=
f_\alpha(q^A, \omega_j(q, \dot q, s)).
\end{eqnarray}
Eq. (\ref{14}) follows from Eqs. (\ref{12}), (\ref{1}). Eq. (\ref{15}) is a consequence of the identities (\ref{1}),
(\ref{11}). Eq. (\ref{15}) will be crucial for discussion of local symmetries
in the next section. At last, Eq. (\ref{16}) is a consequence of Eqs. (\ref{15}), (\ref{1}).

Following to the standard prescription [7, 8], let us construct the Hamiltonian formulation
for $\tilde L$. By using of Eqs. (\ref{15}), (\ref{16}), one finds the conjugate momenta for $q^A, s^a$
\begin{eqnarray}\label{17}
\tilde p_i=\frac{\partial\tilde L}{\partial\dot q^i}=\omega_i(q^A, \dot q^A, s^a), \qquad
\tilde p_\alpha=\frac{\partial\tilde L}{\partial\dot q^\alpha}=f_\alpha(q^A, \omega_j), \cr
\pi_a=\frac{\partial\tilde L}{\partial\dot s^a}=0. \qquad \qquad \qquad \quad
\end{eqnarray}
Due to the identities (\ref{11}), these expressions can be rewritten in the equivalent form
\begin{eqnarray}\label{18}
\dot q^i=v^i(q^A, \tilde p_j, \dot q^\alpha)+s^a\frac{\partial T_a(q^A, \tilde p_j)}{\partial\tilde p_i}, ~
\tilde p_\alpha-f_\alpha(q^A, \tilde p_j)=0, ~ \pi_a=0.
\end{eqnarray}
Thus the velocities $\dot q^i$ have been
determined. There are presented trivial constraints $\pi_a=0$, in a number equal to the number of all the
higher stage constraints of the initial formulation, as well as all the primary constraints $\Phi_\alpha=0$
of the initial theory. Using the definition (\ref{6}), one obtains the Hamiltonian
$\tilde H_0$ $=$ $H_0+s^aT_a$, so the complete Hamiltonian for $\tilde L$ is given by the expression
\begin{eqnarray}\label{19}
\tilde H%(q^A, \tilde p_A, s^a, \pi_a, v^\alpha, v^a)
=H_0(q^A, \tilde p_j)+
s^aT_a(q^A, \tilde p_j)
+v^\alpha\Phi_\alpha(q^A, \tilde p_B)+v^a\pi_a, \quad
\end{eqnarray}
where $v^\alpha, v^a$ are multipliers corresponding to the primary constraints.
Note that, if one discards the constraints $\pi_a=0$, $\tilde H$ coincides with the extended Hamiltonian for $L$ after
identification of configuration space variables $s^a$ with the Lagrangian
multipliers for higher stage constraints of the original formulation.

Further, preservation in time of the
primary constraints $\pi_a$ implies the equations $T_a=0$. Hence all the higher stage constraints of the initial
formulation appear now as the secondary constraints. Preservation in time of the primary constraints $\Phi_\alpha$
leads to the equations $\{\Phi_\alpha, \tilde H\}$ $=$
$\{\Phi_\alpha, H_0\}$ $+$ $\{\Phi_\alpha, \Phi_\beta\}v^\beta$ $+$ $\{\Phi_\alpha, T_b\}s^b$ $=$ $0$.
In turn, preservation of the secondary constraints $T_a$ leads to the similar equations
$\{T_a, \tilde H\}$ $=$
$\{T_a, H_0\}$ $+$ $\{T_a, \Phi_\beta\}v^\beta$ $+$ $\{T_a, T_b\}s^b$ $=$ $0$. To continue the analysis, it is
convenient to unify them as follows:
\begin{eqnarray}\label{20}
\{G_I, H_0\}+\{G_I, G_J\}S^J=0.
\end{eqnarray}
Here $G_I$ are all the constraints of the initial formulation and it was denoted $S^J\equiv(v^\alpha, s^a)$.
Using the matrix (\ref{7.2}), the system (\ref{20}) can be rewritten in the equivalent form
\begin{eqnarray}\label{21}
\{G_{I_1}, H_0\}+O(G_I)=0,
\end{eqnarray}
\begin{eqnarray}\label{22}
\{G_{I_2}, H_0\}+\{G_{I_2}, G_J\}S^J=O(G_I).
\end{eqnarray}
Eq. (\ref{21}) does not contain any new information, since the first class constraints commute with the Hamiltonian,
see Eq. (\ref{8}). So, let us analyze the system (\ref{22}). First, one notes that due to the rank condition
$\left.rank\{ G_{I_2}, G_J\}\right|_{G_I}$ $=$ $[I_2]=max$, exactly $[I_2]$ variables among $S^I$ can be found from
the equations. According to the Dirac prescription, one needs to find maximal number of $v^\alpha$. To make this,
let us restore $v$-dependence in Eq. (\ref{22}):
$\{G_{I_2}, \Phi_\alpha\}v^\alpha$ $+$ $\{G_{I_2}, H_0\}+\{G_{I_2}, T_b\}s^b$ $=$ $0$.
Since the matrix $\{G_{I_2}, \Phi_\alpha\}$ is the same as in the initial formulation, from these equations one
determines some group of variables $v^{\alpha_2}$ through the remaining variables $v^{\alpha_1}$, where $[\alpha_2]$
is number of primary second-class constraints among $\Phi_\alpha$. After substitution of the result into the
remaining equations of the system (\ref{22}), the latter acquires the form
\begin{eqnarray}\label{23}
v^{\alpha_2}=v^{\alpha_2}(q, p, s^a, v^{\alpha_1}), \qquad
Q_{a_2 b}(q, p)s^b+P_{a_2}(q, p)=0,
\end{eqnarray}
where $[a_2]$ is the number of higher-stage second class constraints of the initial theory.
It must be $P\approx 0$, since for $s^b=0$ the system (\ref{22}) is a subsystem of
(\ref{7.1}), but the latter vanish after substitution of the multipliers determined during the procedure,
see discussion after Eq. (\ref{7.1}). Besides, one notes that $rank Q=[a_2]=max$. Actually, suppose that
$rank Q=[a']<[a_2]$. Then from Eq. (\ref{22}) only $[\alpha_2]+[a']<[I_2]$
variables among $S^I$ can be determined, in contradiction with the conclusion made before. In resume, the
system (\ref{20}) for determining the second-stage  and the third-stage constraints and multipliers is
equivalent to the following one
\begin{eqnarray}\label{24}
v^{\alpha_2}=v^{\alpha_2}(q, p, s^{a_1}, v^{\alpha_1}),
\end{eqnarray}
\begin{eqnarray}\label{25}
s^{a_2}=Q^{a_2}{}_{b_1}(q, p)s^{b_1}.
\end{eqnarray}
Conservation in time of the constraints (\ref{25}) does not produce new constraints, giving equations for
determining the multipliers
\begin{eqnarray}\label{25.1}
v^{a_2}=\{ Q^{a_2}{}_{b_1}(q, p)s^{b_1}, \tilde H\},
\end{eqnarray}
The Dirac procedure for $\tilde L$ stops on this stage. All the constraints of the theory have been revealed
after completing the third stage.

Now we are ready to compare the theories $\tilde L$ and $L$. Dynamics of the theory $\tilde L$ is governed by the
Hamiltonian equations
\begin{eqnarray}\label{26}
\dot q^A=\{q^A, H\}+s^a\{q^A, T_a\}, \qquad \dot{\tilde p}_A=\{\tilde p_A, H\}+s^a\{\tilde p_A, T_a\}, \cr
\dot s^a=v^a, \qquad \qquad \qquad \qquad \qquad \dot\pi_a=0, \qquad \qquad \qquad \qquad \quad
\end{eqnarray}
as well as by the constraints
\begin{eqnarray}\label{27}
\Phi_\alpha=0, \qquad T_a=0,
\end{eqnarray}
\begin{eqnarray}\label{28}
\pi_{a_1}=0,
\end{eqnarray}
\begin{eqnarray}\label{29}
\pi_{a_2}=0, \qquad s^{a_2}=Q^{a_2}{}_{b_1}(q, p)s^{b_1}.
\end{eqnarray}
Here $H$ is complete Hamiltonian of the initial theory (\ref{5}), and the Poisson bracket is defined
on the phase space $q^A, s^a, p_A, \pi_a$. The constraints $\pi_{a_1}=0$ can be replaced by the combinations
$\pi_{a_1}-\pi_{a_2}Q^{a_2}{}_{a_1}(q, p)=0$, the latter represent first class subset.
Let us make partial fixation of a gauge by imposing the equations $s^{a_1}=0$ as a gauge conditions for the
subset. Then $(s^a, \pi_a)$-sector of the theory disappears, whereas the equations (\ref{26}), (\ref{27}) coincide
exactly with those of the initial
theory\footnote{In more rigorous treatment,
one writes Dirac bracket corresponding to the equations $\pi_{a_1}-\pi_{a_2}Q^{a_2}{}_{a_1}=0$, $s^{a_1}=0$,
and to the second class constraints (\ref{29}). After that, the equations used in construction of the Dirac bracket
can be used as strong equalities. For the case, they reduce to the equations $s^a=0, \pi_a=0$. For the remaining phase-space
variables $q^A, p_A$, the Dirac bracket coincides with the Poisson one.} $L$.
Let us remind that $\tilde L$
has been constructed in some vicinity of the point $s^a=0$. The gauge $s^{a_1}=0$ implies $s^a=0$ due to the homogeneity
of Eq. (\ref{25}). It guarantees a self consistency of the construction. Thus $L$ represents one of the
gauges for $\tilde L$, which proves equivalence of the two formulations.

\section{Local symmetries of the extended Lagrangian}

Since the initial Lagrangian is one of gauges for $\tilde L$, physical system under consideration can
be equally analyzed by using of the extended Lagrangian. In contrast to $L$, the extended Lagrangian
contains the higher-stage constraints $T_a$ of $L$ in the manifest form, see Eq. (\ref{13}).
Moreover, while $T_a$ appear as the secondary constraints of the formulation $\tilde L$, they are also
presented in the manifest form in the complete Hamiltonian $\tilde H$.
Here we demonstrate one of consequences of this
property: all the infinitesimal local symmetries of $\tilde L$ can be found in closed form.

According to the analysis made in the previous section, the primary constraints of the extended formulation are
$\Phi_\alpha=0$, $\pi_a=0$. Among $\Phi_\alpha=0$ there are presented first class constraints, in a
number equal to the number of primary first class constraints of $L$. Among
$\pi_a=0$, we have find the first class constraints $\pi_{a_1}-\pi_{a_2}Q^{a_2}{}_{a_1}(q, p)=0$, in
a number equal to the number of all the higher-stage first class constraints of $L$.
Thus the number of primary first class constraints of $\tilde L$ coincide with the number $[I_1]$
of all the first class constraints of $L$. Hence one expects $[I_1]$ local symmetries presented in the
formulation $\tilde L$.
Now we demonstrate that the action $S_{\tilde L}=\int d\tau\tilde L$ is invariant (modulo to a surface term) under
the following infinitesimal transformations:
\begin{eqnarray}\label{30}
\delta_{I_1} q^A=\epsilon^{I_1}\left.\{q^A, G_{I_1}(q^A, p_B)\}
\right|_{p_i\rightarrow\omega_i(q, \dot q, s), p_\alpha\rightarrow f_\alpha(q, \omega(q, \dot q, s))}, \qquad \qquad ~ ~\cr
\delta_{I_1} s^a= \qquad \qquad \qquad \qquad \qquad \qquad \qquad \qquad \qquad \qquad \qquad \quad \cr
\left.\left[\dot\epsilon^{I_1}K_{I_1}{}^a+\epsilon^{I_1}\left( b_{I_1}{}^a+s^bc_{I_1 b}{}^a+
\dot q^\beta c_{I_1 \beta}{}^a\right)\right]
\right|_{p_i\rightarrow\omega_i(q, \dot q, s), p_\alpha\rightarrow f_\alpha(q, \omega(q, \dot q, s))}.
\end{eqnarray}
Here $\epsilon^{I_1}(\tau)$, $I_1=1, 2, \ldots , [I_1]$ are the local parameters, and $K$ is the conversion
matrix, see Eq. (\ref{7.2}).
Eq. (\ref{30}) gives the symmetries of $\tilde L$ in closed form in terms of the first class
constraints $G_{I_1}$ of the initial formulation. One should note that the transformations of $q^A$
represent Lagrangian version of canonical transformations with the generators being $G_{I_1}$.

In the subsequent computations we omit all the terms which are total derivatives. Besides,
the notation $\left. A\right|$ implies the substitution indicated in Eq. (\ref{30}).

To make a proof, it is convenient to represent the extended Lagrangian (\ref{13}) in terms of the initial
Hamiltonian $H_0$, instead of the initial Lagrangian $L$. With help of Eq. (\ref{6}) one writes
\begin{eqnarray}\label{31}
\tilde L(q^A, \dot q^A, s^a)=
\omega_i\dot q^i+f_\alpha(q^A, \omega_j)\dot q^\alpha-
H_0(q^A, \omega_j)-s^aT_a(q^A, \omega_j),
\end{eqnarray}
where the functions $\omega_i(q, \dot q, s)$, $f_\alpha(q, \omega)$ are defined by Eqs. (\ref{10}), (\ref{4}).
Using the identity (\ref{15}), variation of this expression under the transformation (\ref{30}) can be
presented in the form
\begin{eqnarray}\label{32}
\delta\tilde L=-\dot\omega_i(q, \dot q, s)\left.\frac{\partial G_{I_1}}{\partial p_i}\right|\epsilon^{I_1}
-\dot f_\alpha(q, \omega(q, \dot q, s)\left.\frac{\partial G_{I_1}}{\partial p_\alpha}\right|\epsilon^{I_1}
\qquad \qquad \quad  ~ \cr
-\left. \left(\frac{\partial H_0(q^A, p_j)}{\partial q^A}+
\dot q^\alpha\frac{\partial\Phi_\alpha(q^A, p_B)}{\partial q^A}+
s^a\frac{\partial T_a(q^A, p_j)}{\partial q^A}\right)\right|\left.\{q^A, G_{I_1}\}\right|\epsilon^{I_1} \cr
-\delta_{I_1}s^aT_a(q^A, \omega_j).\qquad \qquad \qquad \qquad \qquad \qquad \qquad \qquad \qquad \qquad \quad
\end{eqnarray}
To see that $\delta\tilde L$ is total derivative, we add the following zero
\begin{eqnarray}\label{33}
0\equiv\left.\left[\left.\frac{\partial\tilde L}{\partial\omega_i}\right|_{\omega_i}\{p_i, G_{I_1}\}\right.\right.
\qquad \qquad \qquad \qquad \qquad \qquad \qquad \quad ~\cr
\left.\left.
-\left(\frac{\partial H_0}{\partial p_\beta}+
\dot q^\alpha\frac{\partial\Phi_\alpha}{\partial p_\beta}+
s^a\frac{\partial T_a}{\partial p_\beta}\right)\{p_\beta, G_{I_1}\}+
\dot q^\alpha\{p_\alpha, G_{I_1}\}\right]\right|\epsilon^{I_1},
\end{eqnarray}
to r.h.s. of Eq. (\ref{32}). It gives the expression
\begin{eqnarray}\label{34}
\delta\tilde L=
\left.\left[\dot\epsilon^{I_1}G_{I_1}-\epsilon^{I_1}\left(\{H_0, G_{I_1}\}+
\dot q^\alpha\{\Phi_\alpha, G_{I_1}\}+s^a\{T_a, G_{I_1}\}\right)\right]\right| \cr
-\delta_{I_1}s^aT_a(q^A, \omega_j)= \qquad \qquad \qquad \qquad \qquad \qquad \qquad \qquad \qquad \cr
\left.\left[\dot\epsilon^{I_1}G_{I_1}+\epsilon^{I_1}\left(b_{I_1}{}^I+
\dot q^\alpha c_{I_1 \alpha}{}^I+s^bc_{I_1 b}{}^I\right)G_I\right]\right|-\delta_{I_1}s^aT_a(q^A, \omega_j),
\end{eqnarray}
where $b, c$ are coefficient functions  of the constraint algebra (\ref{8}).
Using the equalities $\left. G_{I}\right|=(0, ~  T_a(q^A, \omega_j))$,
$\left. G_{I_1}\right|=K_{I_1}{}^a T_a(q^A, \omega_j)$, one finally obtains
\begin{eqnarray}\label{35}
\delta\tilde L= \qquad \qquad \qquad \qquad \qquad \qquad \qquad \cr
\left.\left[\dot\epsilon^{I_1}K_{I_1}{}^a+\epsilon^{I_1}\left(b_{I_1}{}^a+
\dot q^\alpha c_{I_1 \alpha}{}^a+s^bc_{I_1 b}{}^a\right)-\delta_{I_1}s^a\right]\right|_
{p_i\rightarrow\omega_i}T_a.
\end{eqnarray}
Then the variation of $s^a$ given in Eq. (\ref{30}) implies $\delta\tilde L=div$, as it has been stated.

\section{Conclusion}
In this work we have presented a relatively simple way for finding the
local symmetries in a singular theory of a general form.
Instead of looking for the symmetries of initial Lagrangian, one can construct an equivalent
Lagrangian $\tilde L$ given by Eq. (\ref{13}), the latter implies at most third-stage constraints in
the Hamiltonian formulation\footnote{In the recent work [9] it was demonstrated that the primary
constraints, while are convenient, turn out to be not necessary for the Hamiltonization procedure. So, one
can said that for any theory there exists a formulation with secondary and tertiary constraints.}.
Due to special structure of $\tilde L$ (all the higher-stage constraints $T_a$ of the original
formulation enter into $\tilde L$ in a manifest form, see the last term in Eq. (\ref{13})),
local symmetries of $\tilde L$ can be immediately written according to Eq. (\ref{30}).
The latter gives the symmetries in terms of the first class
constraints $G_{I_1}$ of the initial formulation, moreover, transformations of $q^A$
represent Lagrangian version of canonical transformations with the generators being $G_{I_1}$.
In contrast to a situation with symmetries of $L$ [2-5], the transformations (\ref{30}) do not involve
the second class constraints.

The extended formulation can be appropriate tool for development of a general formalism for conversion
of second class constraints into the first class ones according to the ideas of the work [10]. To
apply the method proposed in [10], it is desirable to have a formulation with some configuration
space variables entering into the Lagrangian without derivatives. It is exactly
what happens in the extended formulation.

To conclude with, we discuss properties of the extended formulation for some particular cases of the
original gauge algebra (\ref{8}).

Suppose that all the original constraints $G_I$ are first class. It implies the extended formulation
with at most secondary constraints. One obtains the primary constraints $\Phi_\alpha=0$, $\pi_a=0$ and
the secondary constraints $T_a=0$, all of them being the first class. An appropriate gauge for $\pi_a=0$
is $s^a=0$. For the case, Eq. (\ref{30}) reduces to the result obtained in [6].

Suppose that all the original constraints $G_I$ are second class (that is there are no of local
symmetries in the theory). It implies the extended formulation
with at most third-stage constraints, all of them being the second class: $\Phi_\alpha=0$, $T_a=0$, $\pi_a=0$,
$s^a=0$.

Suppose that the original $L$ represents a formulation with at most second-stage first and second class
constraints. It implies the extended formulation with at most third-stage constraints. Nevertheless,
namely for the extended formulation the local symmetries can be find in a manifest
form according to Eq. (\ref{30}).

\section{Acknowledgments}
Author would like to thank the Brazilian foundations CNPq (Conselho Nacional de Desenvolvimento
Científico  e Tecnológico - Brasil) and FAPERJ for financial support.

\end{document}